\documentclass[aps,prl,preprintnumbers,superscriptaddress,showpacs,showkeys,floatfix,amsmath,amssymb,amsfonts,twocolumn]{revtex4-2}

\usepackage{graphicx}
\usepackage{multirow}
\usepackage{xparse} 
\usepackage{float}
\usepackage{xcolor}
\usepackage{makecell}
\usepackage{enumitem}

\NewDocumentCommand\Nf{mgg}{N\textsubscript{f}=#1\IfNoValueTF{#2}{}{+#2}\IfNoValueTF{#3}{}{+#3}}
\NewDocumentCommand\vol{mg}{#1\textsuperscript{3}\IfNoValueTF{#2}{}{×#2}}

\newcommand{\tins}{t_\mathrm{ins}}
\newcommand{\xins}{x_\mathrm{ins}}
\newcommand{\vxins}{\vec{x}_\mathrm{ins}}
\usepackage[normalem]{ulem}

\usepackage{hyperref}
\usepackage{slashed}

\begin{document}

\title{Strangeness of nucleons from \Nf{2}{1}{1} lattice QCD}

\author{Constantia Alexandrou} \affiliation{Computation-based Science and Technology Research Center, The Cyprus Institute, 20 Kavafi Str., Nicosia 2121, Cyprus}\affiliation{Department of Physics, University of Cyprus, P.O. Box 20537, 1678 Nicosia, Cyprus}
\author{Simone Bacchio} \affiliation{Computation-based Science and Technology Research Center, The Cyprus Institute, 20 Kavafi Str., Nicosia 2121, Cyprus}
\author{Mathis Bode} \affiliation{J\"{u}lich Supercomputing Centre, Forschungzentrum Jülich, D-52425 Jülich, Germany}
\author{Jacob Finkenrath} \affiliation{Department of Physics, Bergische Universit\"{a}t Wuppertal,
Gaußstr. 20, 42119 Wuppertal, Germany}
\author{Andreas Herten} \affiliation{J\"{u}lich Supercomputing Centre, Forschungzentrum Jülich, D-52425 Jülich, Germany}
\author{Christos Iona} \affiliation{Computation-based Science and Technology Research Center, The Cyprus Institute, 20 Kavafi Str., Nicosia 2121, Cyprus}
\author{Giannis Koutsou} \affiliation{Computation-based Science and Technology Research Center, The Cyprus Institute, 20 Kavafi Str., Nicosia 2121, Cyprus}
\author{Ferenc Pittler} \affiliation{Computation-based Science and Technology Research Center, The Cyprus Institute, 20 Kavafi Str., Nicosia 2121, Cyprus}
\author{Bhavna Prasad} \affiliation{Computation-based Science and Technology Research Center, The Cyprus Institute, 20 Kavafi Str., Nicosia 2121, Cyprus}
\author{Gregoris Spanoudes}\affiliation{Department of Physics, University of Cyprus, P.O. Box 20537, 1678 Nicosia, Cyprus}

\date{\today}

\begin{abstract}
	We present the strange electromagnetic form factors of the nucleon
	using lattice QCD simulations with degenerate light, a strange, and a
	charm quark in the sea with masses tuned to their physical values.
	For the first time, the strange electromagnetic form factors are
	computed at the continuum limit using only ensembles simulated with
	physical quark masses, eliminating the need for chiral extrapolations
	and their associated systematic uncertainty. We obtain the momentum
	transfer dependence of the form factors using the $z$-expansion and
	provide the strange electric and magnetic radii, as well as the
	strange magnetic moment.  When combining our statistical errors and
	systematic uncertainties stemming from the momentum transfer
	dependence fit, our errors are an order of magnitude smaller than
	those associated with experimental determinations of the strange
	electromagnetic form factor.

\end{abstract}

\maketitle

\paragraph{Introduction.}

A key nonperturbative feature of Quantum Chromodynamics (QCD) is the
creation of virtual quark and antiquark pairs in the QCD vacuum. These
sea-quark fluctuations are important for understanding the internal
structure of hadrons. Nucleon electromagnetic form factors offer a
clear and systematic way to study this structure, as they reveal how
charge and magnetization are distributed inside the nucleon in the
non-relativistic limit. Sea quark effects are especially interesting
for the strange quark. As the lightest non-valence quark its
contribution arises purely from vacuum fluctuations, giving direct
insight into nonperturbative QCD vacuum dynamics. The strange
electromagnetic form factors can be measured experimentally by
examining parity-violating asymmetries in elastic polarized
electron-unpolarized nucleon scattering. These asymmetries result from
the mixing of electromagnetic and weak neutral-current interactions. A
number of experiments over the years, such as
SAMPLE~\cite{SAMPLE:2003wwa, Beise:2004py},
A4~\cite{A4:2004gdl,Maas:2004dh,Baunack:2009gy},
HAPPEX~\cite{HAPPEX:2005qax,HAPPEX:2005zgj,HAPPEX:2006oqy,HAPPEX:2011xlw},
and G0~\cite{G0:2005chy,G0:2009wvv}, have provided results for the
strange magnetic moment and electric and magnetic radii. Although
indicating a non-zero value, these results have large errors and thus
do not exclude the zero value for the strange magnetic moment and the
electric and magnetic radii.

A high-precision determination within lattice QCD is thus essential to
disentangle these small contributions from the dominant up- and
down-quark components. In particular, reliable lattice calculations of
the strange form factors provide critical theoretical input for the
extraction of the proton weak charge from parity-violating electron
scattering. This program has been pursued with high precision by the
$Q_{\rm weak}$ experiment at Jefferson Lab~\cite{Qweak:2018tjf}, where
control over hadronic structure effects, including strange quark
contributions, is necessary to match the experimental accuracy and to
enable stringent tests of the Standard Model. Although there have been
lattice QCD results in the past for the strange electromagnetic form
factors~\cite{Djukanovic:2019jtp,Alexandrou:2019olr,Sufian:2016pex,Green:2015wqa},
the analysis has either been done at heavier than physical quark
masses or includes ensembles at non-physical quark mass values in
order to obtain results at the physical pion mass through chiral
extrapolations. In this work, we provide the first calculation of the
strange electromagnetic form factors of the nucleon within lattice QCD
using only ensembles with quark masses tuned to their physical values
(physical point). Specifically, we use four ensembles of
clover-improved twisted mass fermions with two degenerate light, a
strange, and a charm quark (\Nf{2}{1}{1}) at four different lattice
spacings enabling us to take a controlled continuum limit directly at
the physical pion mass point.

\paragraph{Matrix element of the nucleon electromagnetic current.}

The electromagnetic form factors are given in terms of the matrix
element of the electromagnetic current with nucleon states in
Minkowski space,
\begin{gather}
	\langle N(p',s') \vert j_\mu \vert N(p,s) \rangle = \sqrt{\frac{m_N^2}{E_N(\vec{p}\,') E_N(\vec{p})}} \times \label{eq:me} \\
	\bar{u}_N(p',s') \left[ \gamma_\mu F_1(q^2) + \frac{i \sigma_{\mu\nu} q^\nu}{2 m_N} F_2(q^2) \right] u_N(p,s),\, \nonumber
\end{gather}
where $N(p,s)$ is the nucleon with initial (final) momentum $p$ ($p'$)
and spin $s$ ($s'$), with energy $E_N(\vec{p})$ ($E_N(\vec{p}\,') $)
and mass $m_N$, $u_N$ is the nucleon spinor, $j_\mu$ is the vector
current, and $q^2{\equiv}q_\mu q^\mu$ is the momentum transfer squared
with $q_\mu{=}(p_\mu'-p_\mu)$. We work in the flavor isospin symmetric
limit, where up- and down-quarks are degenerate. The local vector
current for the strange contribution to the nucleon is given by, $
	j_\mu = e_s j^s_\mu = e_s \bar{\psi}_s \gamma_\mu \psi_s$, where $s$
denotes the strange quark flavor. The strange electric charge,
$e_s=-1/3$, is not included in results presented in this study. We
will quote results for the electric and magnetic Sachs form factors,
$G^s_E(q^2)$ and $G^s_M(q^2)$ respectively, which are given in terms
of the Dirac ($F^s_1$) and Pauli ($F^s_2$) form factors via
\begin{gather}
	G^s_E(q^2) = F^s_1(q^2) + \frac{q^2}{4m_N^2} F^s_2(q^2),\,\,\, \\
	G^s_M(q^2) = F^s_1(q^2) + F^s_2(q^2)\,.\label{eq:sachs}
\end{gather}
The electric and magnetic
mean-squared radii are defined as the slope of the
corresponding Sachs form factor as $q^2\rightarrow 0$, namely
$\langle r_X^2 \rangle^s = -6
	\frac{\partial G_X^s(q^2)}{\partial q^2} \Big \vert_{q^2=0},\,
	\label{eq:radius}$
with $X=E,M$.  In the next section, we briefly summarize our lattice setup,
extraction of form factors from the lattice data and our final
results. Additional details on the lattice setup and analysis  are presented in the accompanying
paper~\cite{Alexandrou:prdstrange}.

\paragraph{Lattice setup.}

\begin{figure}
	\centering
	\includegraphics[width=0.25\textwidth]{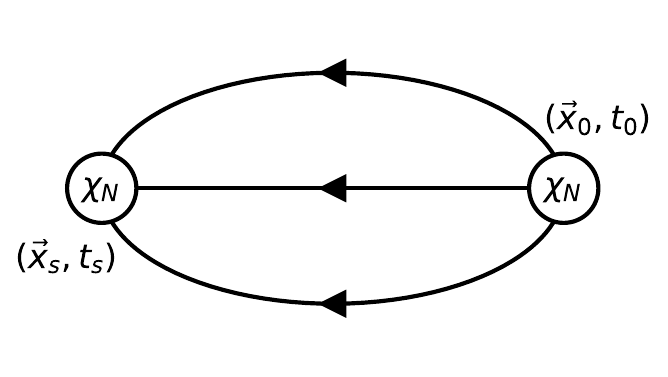}
	\includegraphics[width=0.25\textwidth]{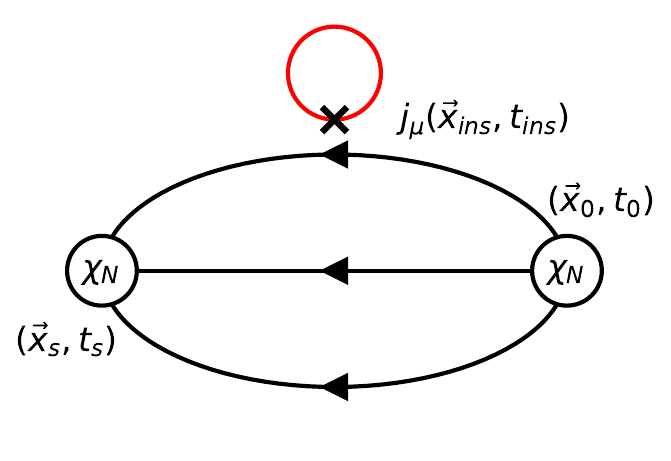}
	\caption{Nucleon two-point function (top) and disconnected nucleon three-point function (bottom).}
	\label{fig:diagrams}
\end{figure}

In order to access the desired nucleon matrix element in lattice QCD, we compute  two- and three-point correlation
functions, as shown in Fig.~\ref{fig:diagrams}. The Euclidean space-time definition in momentum space of the two-point function is given by
\begin{gather}
	C(\Gamma_0,\vec{p};t_s,t_0) = \sum_{\vec{x}_s}
	e^{{-}i (\vec{x}_s{-}\vec{x}_0) \cdot \vec{p}}{\times} \\
	\mathrm{Tr} \left[ \Gamma_0 {\langle}\chi_N(t_s,\vec{x}_s) \bar{\chi}_N(t_0,\vec{x}_0) {\rangle} \right] \,, \label{eq:twop}
\end{gather}
where
$\chi_N(\vec{x},t)=\epsilon^{abc}u^a(x)[u^{b\intercal}(x)\mathcal{C}\gamma_5d^c(x)]\,,$
is the nucleon interpolating field, $\mathcal{C}{=}\gamma_0 \gamma_2$
is the charge conjugation matrix and $Q^2=-q^2$ is the Euclidean
momentum transfer squared. The three-point function is given by
\begin{align}
	C_\mu(\Gamma_\nu,\vec{q},\vec{p}\,';t_s,\tins,t_0) {=} &
	\sum_{\vxins,\vec{x}_s}  e^{i (\vxins {-} \vec{x}_0)  \cdot \vec{q}}  e^{-i(\vec{x}_s {-} \vec{x}_0)\cdot \vec{p}\,'} {\times} \nonumber \\
	\mathrm{Tr} [ \Gamma_\nu \langle \chi_N(t_s,\vec{x}_s) & j_\mu(\tins,\vxins) \bar{\chi}_N(t_0,\vec{x}_0) \rangle].
	\label{eq:thrp}
\end{align}
The initial lattice site  $x_0$ where the nucleon is created is referred to as the
\textit{source}, the lattice site where the current couples to a quark  $\xins$ as the \textit{insertion}, and the site where the nucleon is annihilated  $x_s$ as
the \textit{sink}. $\Gamma_\nu$ is a projector acting on Dirac
indices, with $\Gamma_0 {=} \frac{1}{2}(1{+}\gamma_0)$ yielding the
unpolarized and $\Gamma_k{=}\Gamma_0 i \gamma_5 \gamma_k$ the
polarized matrix elements. Without loss of generality we will take
$t_s$ and $\tins$ relative to the source time $t_0$ in what follows.

In order to extract the ground-state nucleon matrix element and cancel overlaps of the interpolating operator with the nucleon states, an
optimized ratio composed of two- and three-point functions is used, given by
\begin{gather}
	R_{\mu}(\Gamma_{\nu},\vec{p},\vec{p}\,';t_s,\tins) = \frac{C_{\mu}(\Gamma_{\nu},\vec{p},\vec{p}\,';t_s,\tins\
		)}{C(\Gamma_0,\vec{p}\,';t_s)} \times\\
	\sqrt{\frac{C(\Gamma_0,\vec{p};t_s-\tins) C(\Gamma_0,\vec{p}\,';\tins) C(\Gamma_0,\vec{p}\,';t_s)}{C(\Gamma_0,\vec{p}\,';t_s-\tins) C(\Gamma_0,\vec{p};\tins) C(\Gamma_0,\vec{p};t_s)}}.
	\label{eq:full_ratio}
\end{gather}
In the limit of large time separations $(t_s-\tins) \gg$ and $\tins
	\gg$, the ratio in Eq.~(\ref{eq:full_ratio}) converges to the nucleon
ground state matrix element, i.e.
\begin{gather}
	R_{\mu}(\Gamma_{\nu},\vec{p},\vec{p}\,';t_s,\tins)\xrightarrow[(t_s-\tins)
		\gg]{\tins\gg}\Pi_{\mu}(\Gamma_{\nu},\vec{p},\vec{p}\,').
	\label{eq:ratio_limit}
\end{gather}

\begin{table}[h]
	\caption{Parameters of the four \Nf{2}{1}{1} ensembles analyzed in
		this work. From the leftmost to rightmost columns, we provide
		the name of the ensemble, the lattice volume, $\beta=6/g^2$ with
		$g$ the bare coupling constant, the lattice spacing, the pion
		mass, and the value of $m_\pi L$. Lattice spacings and pion
		masses are taken from
		Ref.~\cite{ExtendedTwistedMass:2022jpw,ExtendedTwistedMass:2024nyi}.}
	\label{tab:ens}
	\centering
	\begin{tabular}{cccccccc}
		\hline\hline
		Ensemble               & $(\frac{L}{a})^3{\times}(\frac{T}{a})$ & $\beta$ & \makecell[c]{$a$                   \\$[$fm$]$} & \makecell[c]{$m_\pi$\\ $[$MeV$]$}  & $m_\pi L$ \\
		\hline
		\texttt{cB211.072.64}  & $64^3 {\times} 128$                    & 1.778   & 0.07957(13)      & 140.2(2) & 3.62 \\
		\texttt{cC211.060.80}  & $80^3 {\times} 160$                    & 1.836   & 0.06821(13)      & 136.7(2) & 3.78 \\
		\texttt{cD211.054.96}  & $96^3 {\times} 192$                    & 1.900   & 0.05692(12)      & 140.8(2) & 3.90 \\
		\texttt{cE211.044.112} & $112^3 {\times} 224$                   & 1.960   & 0.04892(11)      & 136.5(2) & 3.79 \\
		\hline
	\end{tabular}
\end{table}

We use ensembles simulated with \Nf{2}{1}{1} twisted mass,
clover-improved fermions with quark masses tuned to approximately
their physical values.  This formalism allows for automatic
${\cal{O}}(a)$ improvement without requiring any further improvement
of the operators yielding physical quantities~\cite{Frezzotti:2000nk,
Frezzotti:2003ni}. A summary of the parameters for the ensembles is
provided in Table~\ref{tab:ens}. We employ Gaussian smearing for
increasing the overlap of the nucleon interpolating field with nucleon
ground state~\cite{Gusken:1989qx,Alexandrou:1992ti} and utilize
APE-smeared gauge links~\cite{APE:1987ehd}, with parameters specified
in Ref.~\cite{Alexandrou:2025vto}. The quark field contractions of
Eq.~(\ref{eq:thrp}) give rise to the quark-disconnected three-point
function, where the strange quark loop is given by,
\begin{align}
	L_s(t_\mathrm{ins},\vec{q})=
	\sum_{\vec{x}_\mathrm{ins}}e^{i\vec{q}\cdot\vec{x}_\mathrm{ins}}\mathrm{Tr}[ & D^{-1}_s(x_\mathrm{ins};x_\mathrm{ins})\gamma_\mu].
	\label{eq:gen-one-end}\end{align}
The quark loop is computed using a combination of full dilution in
spin and color~\cite{Alexandrou:2012zz, Alexandrou:2019olr},
hierarchical probing,~\cite{Stathopoulos:2013aci} and deflation of low
modes.

With the quark loop at hand for every time slice $t_{\rm ins}$,
the disconnected three-point function can be computed at every value
of $t_s$ and $t_{\rm ins}$ as well as additional sink momenta by
correlating it with the two-point functions. Details on the number of
source positions ($n_{\rm src}$) used in the computation of the two-
and the three-point functions as well as the number of configurations
($n_{\rm conf}$) available per ensemble are also tabulated in
Table~\ref{tab:disc_range}.

We employ the local vector current to compute the disconnected
three-point functions and therefore, the matrix elements must be renormalized.
We determine the corresponding renormalization factor, $Z_V$, following the procedure described in
our Ref.~\cite{Alexandrou:2025vto}.
In the present work, we extend the calculation of~\cite{Alexandrou:2025vto} to include the finer lattice spacing E.
Furthermore, we improve the analysis by performing joint polynomial fits in the renormalization scale across all four lattice spacings.
The final values of $Z_V$ for all ensembles are listed in
Table~\ref{tab:disc_range}.

\paragraph{Extraction of form factors.}

In order to extract the form factors $G_E(Q^2)$ and $G_M(Q^2)$ at each value of $Q^2$, we need to find an appropriate solution to the equations for
$\Pi_\mu(\Gamma_\nu;\vec{p}', \vec{q})$ of Eq.~(\ref{eq:ratio_limit}) arising from varying $\Gamma_\nu$ and
$\mu$ depending on the momenta $\vec{p}'$ and $\vec{q}$. Since we can access additional sink momenta at no additional computational cost, we use
$\vec{p}'=\frac{2\pi}{L}\vec{k}$ for $\vec{k}^2=0$, 1, and 2. In the case of non-zero sink momentum ($\vec{p}'\neq0$), the expressions yielding $G_E(Q^2)$ and $G_M(Q^2)$ cannot be disentangled
and we, therefore, use a Singular Value Decomposition to solve the overconstrained set of
equations that emerge, see Appendix of Ref.~\cite{Alexandrou:prdstrange} for more details.

\begin{table}
	\centering
	\caption{The number of configurations ($n_{\rm conf}$) and the
		number of source positions ($n_{\rm src}$) used for each
		ensemble. We also list the source-sink time separations used
		($t_s$) and the renormalization constants ($Z_V$) taken from
		Ref.~\cite{Alexandrou:2025vto}.}
	\label{tab:disc_range}
	\begin{tabular}{ccccccc}
		\hline\hline Ensemble  & $n_{\rm conf}$ & $n_{\rm src}$ & $t_s/a$                & $Z_V$     \\
		\hline
		\texttt{cB211.072.64}  & 749            & 349           & 10, 12, 14, 16         & 0.7228(5) \\
		\texttt{cC211.060.80}  & 399            & 650           & 12, 14, 16, 18         & 0.7373(5) \\
		\texttt{cD211.054.96}  & 493            & 368           & 14, 16, 18, 20, 22     & 0.7521(4) \\
		\texttt{cE211.044.112} & 501            & 339           & 16, 18, 20, 22, 24, 26 & 0.7638(4) \\
		\hline
	\end{tabular}
\end{table}

To extract the nucleon matrix element from $\Pi_\mu(\Gamma_\nu;\vec{p}', \vec{q})$, we perform so-called plateau
fits, defined as a constant fit in $\tins$ to the optimized ratio defined in Eq.~(\ref{eq:full_ratio}) for a given $t_s$. We combine
fits for multiple $t_s$ by taking the weighted average of results from plateau fits using $t_s \in [0.8,1.28] \;\rm fm$, where convergence is
observed. For more details on the fitting procedure see Ref.~\cite{Alexandrou:prdstrange}.

\paragraph{Results for form factors.}

In order to obtain the strange electric and magnetic radii, and the magnetic moment, we  parameterize the $Q^2$ dependence of the form factors. 
We adopt a one-step procedure, where we  simultaneously fit for the $Q^2$-dependence  of the form factors computed using each ensemble and for taking the continuum limit. 
This is done by  incorporating the lattice spacing dependence in the fit parameters that enter the functions used for fitting the $Q^2$ dependence. 
Namely, we use for the $a$-dependence of all fit parameters $f(a^2)=f_0+f_1a^2$. We use the following three parameterizations for the $Q^2$ dependence:
\begin{enumerate}[label=\roman*)]
	\item the dipole form
	      \begin{gather}
		      G(Q^2,a^2) = \frac{g(a^2)}{\left[1+\frac{Q^2}{12}\langle r'^2(a^2)\rangle \right]^2};
	      \end{gather}
	\item the Galster-like parameterization
	      \begin{gather}
		      G(Q^2, a^2) = \frac{Q^2 A(a^2)}{4 m_N^2 + Q^2 B(a^2)}\frac{1}{\left(1+\frac{Q^2}{0.71 {\rm GeV}^2}\right)^{2}};
		      \label{eq:Galster-like}
	      \end{gather}
	\item the $z$-expansion parameterization
	      \begin{gather}
		      G(Q^2,a^2) = \sum_{k=0}^{k_{\rm max}} c_k(a^2) z^k(Q^2),
		      \label{eq:zexp}
	      \end{gather}
	      \begin{equation}
		      \text{where }\;\;\; z(Q^2) = \frac{\sqrt{t_{\rm cut} + Q^2} - \sqrt{t_{\rm cut}+t_0} }{
			      \sqrt{t_{\rm cut} + Q^2} + \sqrt{t_{\rm cut}+t_0} }, \nonumber
		      \label{eq:zQ2}
	      \end{equation}
\end{enumerate}
with $t_{\rm cut}=(2m_K)^2$  the kaon production threshold. We take $m_K=486 \rm \;MeV$ and $t_0=0$.
Additionally, we require smooth convergence of the form factor to zero at $Q^2\rightarrow \infty$~\cite{Lee_2015} 
and employ gaussian priors with a width, $w$, which we vary together with the $k_{\max}$ to check for convergence~\cite{Alexandrou:2023qbg,Alexandrou:2025vto}.

\begin{figure}[h]
	\includegraphics[width=\columnwidth]{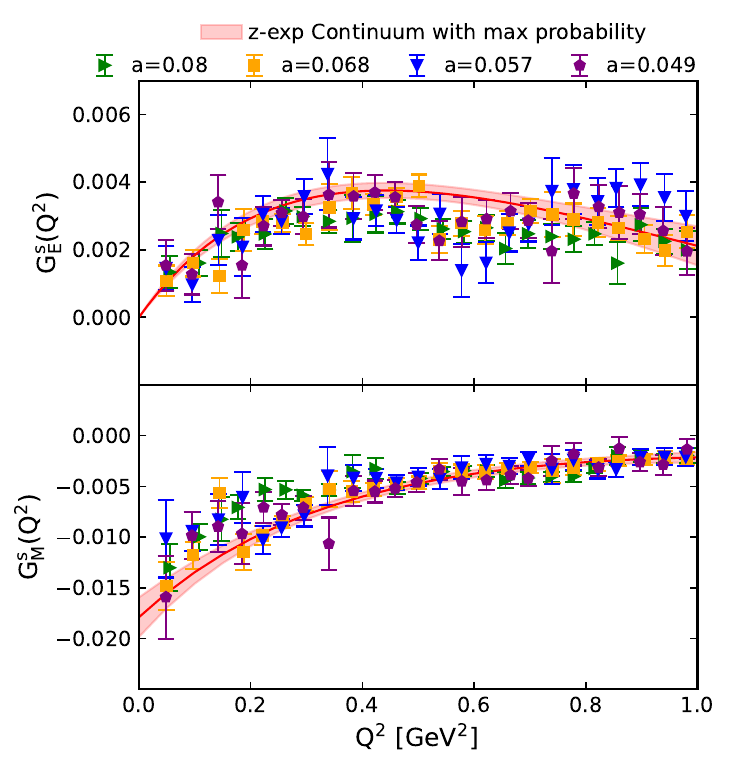}
	\caption{Results for $G^s_E(Q^2)$ (top) and $G^s_M(Q^2)$
		(bottom). The right-pointing green triangles, orange squares,
		downward-pointing blue triangles, and purple pentagons show the
		results of the \texttt{cB211.072.64}, \texttt{cC211.060.80},
		\texttt{cD211.054.96}, and \texttt{cE211.044.112} ensembles,
		respectively. The red band corresponds to the continuum limit
		obtained by fitting to the $z$-expansion.}
	\label{fig:zexp}
\end{figure}

In Fig.~\ref{fig:zexp}, we show results obtained for $G_E^s(Q^2)$ and
$G_M^s(Q^2)$ for all four ensembles analyzed in this work. Since we
have $O(300)$ $Q^2$ values, the results are binned into 25 bins
between $Q^2\in [0,1]$ for better visualization. Access to the
additional momentum frames in the $z$-expansion fit improves the
coefficient constraints, reduces truncation and prior dependence, and
yields more reliable fit results compared to restricting to
$\vec{p}\,'=0$. The red band shows the continuum limit results using
the $z$-expansion for the most probable model obtained using the
Akaike Information Criteria (AIC), a procedure followed in
our previous analyses~\cite{Alexandrou:2023qbg, Alexandrou:2025vto}.

\begin{figure}
	\centering
	\includegraphics[width=\columnwidth]{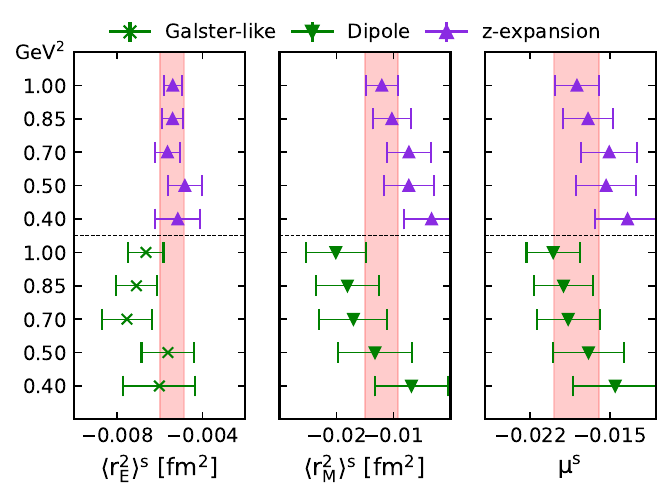}
	\caption{Values of $\langle r^2_E\rangle^s$, $\langle
		r^2_M\rangle^s$, and $\mu^s$ obtained as a result of fitting to
	$G_E^s(Q^2)$ and $G_M^s(Q^2)$ with different values of $Q^2_{\rm
		cut}$. The upward-pointing violet triangles denote results
	from the $z$-expansion fits, the green crosses denote the
	Galster-like fit results, and the downward-pointing green
	triangles denote the results from dipole fits. The red band,
	which runs through vertically, corresponds to the model-averaged
	value obtained using the AIC.}
	\label{fig:Results}
\end{figure}

In Fig.~\ref{fig:Results}, we show the results for the strange
electric ($\langle r^2_E\rangle^s$) and magnetic ($\langle
r^2_M\rangle^s$) radii and the magnetic moment ($\mu^s$) obtained
using the different $Q^2$-parameterizations mentioned above. We
additionally vary the maximum value of $Q^2$ used in the fits,
$Q^2_{\rm cut}$, and model-average using the AIC to obtain the final
result including a statistical and systematic error. This procedure
yields the following values for the strange electric and magnetic
radii and strange magnetic moment:
\begin{gather}
	\langle r^2_E\rangle^s = -0.00545(49)(26) \;\;\rm fm^2, \\
	\langle r^2_M\rangle^s = -0.01212(280)(72) \;\;\rm fm^2,\\
	\mu^s = -0.01792(195)(18).
\end{gather}

\paragraph{Discussion of results.}

\begin{figure}
	\centering
	\includegraphics[width=\columnwidth]{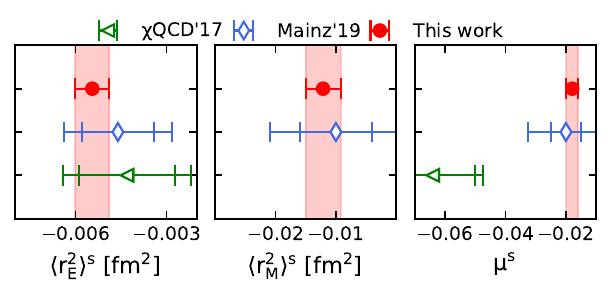}
	\caption{Results for the strange electric and magnetic radii and
		the magnetic moment of the nucleon obtained within this work
		(red circles and vertical band). We compare to previous results
		by $\chi$QCD (left-pointing open triangles)~\cite{Sufian:2016pex}
		and the Mainz group (blue diamonds)~\cite{Djukanovic:2019jtp}.}
	\label{fig:Compare_lattice_results}
\end{figure}

In Fig.~\ref{fig:Compare_lattice_results}, we present a comparison of
our results with other lattice QCD results, all of which include a
continuum extrapolation and chiral extrapolation to physical pion
mass.

The results by $\chi$QCD use four ensembles of \Nf{2}{1} domain wall
sea quarks and overlap valence quarks with $m_{\pi}\in[139-339]\;\rm
MeV$. Their final result is obtained using a continuum and chiral
extrapolation. The Mainz group obtains their
results using six \Nf{2}{1} $\mathcal{O}(a)$-improved Wilson fermion ensembles with
$m_{\pi}\in[200,360] \;\rm MeV$ followed by a continuum and chiral
extrapolation using heavy-baryon chiral perturbation theory.
Our results carry smaller errors arising predominantly due to statistical
uncertainties. As mentioned, our continuum
limit is taken directly at the physical point, thus eliminating the
need for chiral extrapolations, which can lead to a major source of systematic
uncertainty, combined with a large number of statistics for each of
our four ensembles used.

\begin{figure}
	\centering
	\includegraphics[width=\columnwidth]{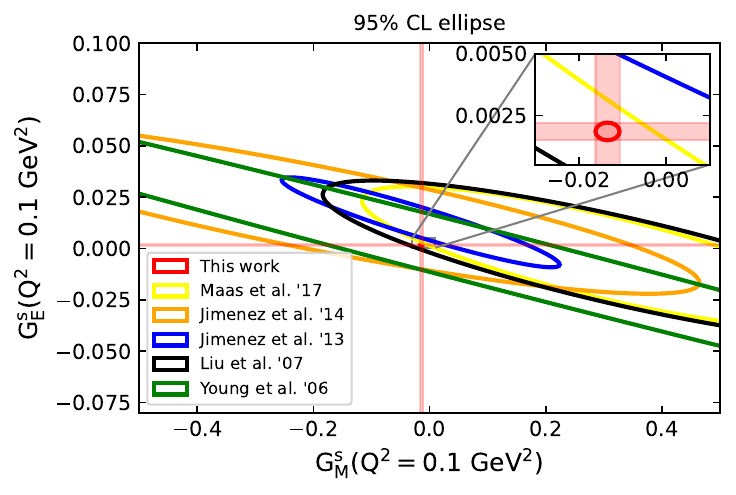}
	\caption{Ellipses showing $95\%$ confidence curves from different
		experimental data. The green ellipse is from
		Ref.~\cite{Young:2006jc}, the orange from
		Ref.~\cite{Gonzalez-Jimenez:2014bia}, the black from
		Ref.~\cite{Liu:2007yi} and the blue from
		Ref.~\cite{Gonzalez-Jimenez:2011qkf}.  The red bands and the red
		ellipse in the inset are from the lattice QCD determination of
		the strange electromagnetic form factors presented in this
		work.}
	\label{fig:Results_exp}
\end{figure}

Having precise results in the continuum limit enables a direct
comparison to experimental results. In Fig.~\ref{fig:Results_exp}, we
show the $95\%$ confidence level contours for $G_E^s$ and $G_M^s$ at
$Q^2=0.1\rm\;GeV^2$ where multiple experimental values are
available. Our results provide much stronger constraints as compared
to the experimental results. This is particularly interesting since it
provides precise input for an upcoming experiment at Mainz, MESA, that
would allow access to the strange electromagnetic form factors through
its parity-violating electron scattering experiments at low momentum
transfer.


\paragraph{Conclusions.}

Precise results on the strange electromagnetic form factors of the
nucleon are presented with lattice systematics
taken into account. Specifically, we use four ensembles with four
different lattice spacings of \Nf{2}{1}{1} twisted mass fermions
simulations performed directly at the physical pion mass value. This
is the first determination of these form factors directly at the
physical point avoiding any uncertainties due to the chiral
extrapolation.
After performing fits to the $Q^2$-dependence of the form factors we
extract precise values for the strange electric and magnetic radii and
magnetic moment.  The results on the electric and magnetic form
factors provide stringent bounds which serve as input for ongoing
analyses and upcoming experiments, as demonstrated in
Fig.~\ref{fig:Results_exp}.

\paragraph{Acknowledgments.}

C.A., S.B., C.I., G.K., F.P., and G.S. acknowledge partial support by
the projects Baryon8, MuonHVP, PulseQCD, DeNuTra, IMAGE-N, HyperON, StrongILA and partonWF (POSTDOC/0524/0001,
EXCELLENCE/0524/0017, EXCELLENCE/0524/0269, EXCELLENCE/0524/0455,
EXCELLENCE/0524/0459, VISION ERC-PATH 2/0524/0001, EXCELLENCE/0524/0001 and VISION ERC/0525/0010, respectively) co-financed by the European
Regional Development Fund and the Republic of Cyprus through the
Research and Innovation Foundation as well as AQTIVATE that received
funding from the European Union’s research and innovation program
under the Marie Sklodowska-Curie Doctoral Networks action, Grant
Agreement No 101072344. C.A acknowledges support by the University of
Cyprus projects ``Nucleon-GPDs'' and ``PDFs-LQCD''. This project
received funding from the European Research Council (ERC) via the
project "LEEX" grant agreement 101170304. Funded by the European
Union. Views and opinions expressed are however those of the author(s)
only and do not necessarily reflect those of the European Union or the
European Research Council Executive Agency (ERCEA). Neither the
European Union nor the ERCEA can be held responsible for them. B.P. is
supported by ENGAGE which received funding from the EU's Horizon 2020
Research and Innovation Programme under the Marie Skłodowska-Curie GA
No. 101034267. This work was supported by grants from the Swiss
National Supercomputing Centre (CSCS) under projects with ids s702 and
s1174. The authors gratefully acknowledge the Gauss Centre for
Supercomputing e.V. (www.gauss-centre.eu) for funding this project by
providing computing time through the John von Neumann Institute for
Computing (NIC) on the GCS Supercomputers JUWELS~\cite{JUWELS} and
JUWELS Booster~\cite{JUWELS-BOOSTER} at J\"ulich Supercomputing Centre
(JSC). The authors acknowledge access to the JUPITER supercomputer,
which is funded by the EuroHPC Joint Undertaking, the German Federal
Ministry of Research, Technology and Space, and the Ministry of
Culture and Science of the German state of North Rhine-Westphalia,
through the JUPITER Research and Early Access Program (JUREAP) as part
of the EuroHPC project application EHPC-EXT-2023E02-052. The authors
also acknowledge the Texas Advanced Computing Center (TACC) at
University of Texas at Austin for providing HPC resources.

\bibliography{refs}

\appendix

\end{document}